\documentclass[12pt]{article}
\newcommand{\be}{\begin{equation}}
\newcommand{\bea}{\begin{eqnarray} \nonumber}
\newcommand{\ee}{\end{equation}}
\newcommand{\eea}{\end{eqnarray}}

\begin{document}

\title {Complexity and  intelligence}

\author{Dipartimento di Fisica, Sezione INFN, SMC and UdRm1 of INFM,\\
Universit\`a di Roma ``La Sapienza'',\\
Piazzale Aldo Moro 2,
I-00185 Rome (Italy)\\
giorgio.parisi@roma1.infn.it}

\maketitle
\abstract{
In this paper I will discuss the properties of the Algorithmic Complexity, presenting the most relevant properties. The 
related concept of logical depth is also introduced. These properties will be used to study the problem of learning from 
example, paying a special attention to machine learning. We introduce the propensity of a machine to learn a rule and we 
use it define the intelligence of a machine.  }

\section{Algorithmic Complexity}

We have already seen in the chapter [\ldots] that Kolmogorov, independently from and more less simultaneously with 
other peoples, introduced the concept of algorithmic complexity of a string of character (or if we prefer of an integer 
number)as the length in bits of the shortest message that is needed to identify such a number.  This look similar to 
Shannon's entropy, but it is deeply different.  Shannon's entropy is the shortest length of the message needed to 
transmit a generic string of characters belonging to a given ensemble, for example a sentence of given length in 
Italian.  Kolmogorov complexity is is the shortest length of the message needed to transmit a given string of 
characters.

In a slightly different setting, that is nowadays more familiar to most of us, we can define the Kolmogorov complexity 
$\Sigma(N)$ of a number $N$ as the length of the shortest computer program that computes such a number (or if we prefer 
prints that number).  This definition of complexity depends on the computer language in which the program is written, so 
that we should indicate in the definition of $\Sigma(N)$ the language we are using. However if we are interested to 
large values of complexity the choice of the language is irrelevant and it contributes in the worst case to constant. 
It is evident that
\be
\Sigma_{Pascal}(N)<\Sigma_{Pascal}(Fortran\ compiler\ written\ in\ Pascal)+\Sigma_{Fortran}(N)
\ee
Indeed if  we know how to compute something in Fortran and we would like to compute it in Pascal, we can just write a 
Fortran compiler in Pascal and use it to execute the Fortran code. In the nutshell we can transmit the Fortran compiler 
written in Pascal 
and the program that computes the message in Fortran. Inverting the argument we get
\be
|\Sigma_{Fortran}(N)-\Sigma_{Pascal}(N)|<const
\ee
where constant does not depend on $N$. So for real complex messages the choice of the language is not important.

This is common experience also in human languages.  If we are going to have a short discussion on the possible evolution of 
the snow conditions in the next days in relation to a trip on a dog-trained sled, it is extremely likely that the 
conversation should be much shorter is some languages and much longer in other ones. On the contrary, if we have to explain 
to somebody the proof of the last Fermat theorem starting from scratch (i.e. from elementary arithmetics), it would 
take years and in the process we can introduce (or invent) the appropriate mathematical language, so that the dependence 
on the natural language we use is rather weak. In other words for messages of small complexity we can profit of the amount 
of information contained in the language, but this help becomes irrelevant in the case of very  complex messages.

It should be clear that in the case of a generic number $N$ of $K$ binary digits, e.g. smaller that $2^{K}$ elementary 
arguments (and also Shannon's theorem) tell us that the complexity is  $K$. The complexity cannot be larger than $K$ 
because we can always write a program that is of length $K$ plus a small constant:

\begin{verbatim}
    write  3178216741756235135123939018297137213617617991101654611661
\end{verbatim}

\noindent The length of this program is the same as the length of the output of the program (apart for the characters lost  in 
writing ``write'') and it rather likely
(I will pay 1000 Euros for a counterexample) that there is no much shorter one \footnote{I beg the reader to forget the 
inefficiency of writing decimal numbers in ASCI.}.
Of course  the program

\begin{verbatim}
    write  11**57
\end{verbatim}

\noindent is much shorter that 

\begin{verbatim}
    write 228761562390246506066453264733492693192365450838991802120171
\end{verbatim}

\noindent although they have the same output.

Let us consider an other example. The program
\begin{verbatim}
    A=2**31-1
    B=5**7
    I=1
    for K=1 to  A-1
        I=mod(I*B,A)
    endfor
\end{verbatim}
generates a sequence of pseudo random numbers that is rather long (about 8 Gbytes) and apart form small variations it is 
extremely likely that it is the shortest program that generate it.
 Of course it is much shorter  of the program (of 8 Gbytes 
of length) that explicitly contains the sequence.

\section{Some properties of complexity and some apparent paradoxes}

Although most of the numbers of $K$ bits have complexity $K$, we have seen that there are notable exceptions. It should 
be clear that to identify these exceptions and to evaluate of the complexity is not a simple task. A number may have a low 
complexity because it is a power, or because it is obtained as the iteration of a simple formula, or just because it 
coincides with the second millions of digits of $\pi$. A systematic search of  these exceptions is not easy.

However one could imagine a simple strategy. We can consider all the programs of length $K$. Each program will stop 
and will write a string (maybe empty) or it will never stop:
A program that never stops is

\begin{verbatim}
    I=1
    do forever
        I=I+1
        if(I<0) then
             write I
             stop
        endif
    enddo
\end{verbatim}
where $I$ is an arbitrary length number.

In this case is trivial to show that the program stops. In other cases is much difficult to decide if the programs 
stops. Let us consider the following example:

\begin{verbatim}
    I=1
    do forever
         I=I+1
         consider all postive integers a,b,c and n, less than I with n>2.
         if(a**n+b**n=c**n) then
             write a,b,c,n
             stop
          endif
      enddo
\end{verbatim}

This program stops if and only if Fermat last theorem is false and we know now that it will never stops. Up to a few 
years ago it was possible that it would stop writing a gigantic number. A  program that stops only if the Goldbach 
conjecture is false quite like will never stop, but we do not know for sure.

There cannot be any computer program that can compute if a computer program stops or not.  Otherwise we would have a 
contradiction.  A computer program would be able to find out those programs who stops, identifies those program that are 
the shortest one and produce the same output: for given length we can sort them in lexicographic order.  If this 
happens, for example wee could identify the first program of length $K$ of this list.  The output of this program has 
complexity $K$, in the same way as all the programs of this list, however we would identify the first program of the 
list transmitting only the number $K$; this can be done using only $ln_{2}(K)$ bits and therefore the complexity of the 
output would be $ln_{2}(K)$, i.e. a number much smaller than $K$ and this is a contradiction.

One could try to find out a loophole in the argument.  Although it is impossible to decide if a program does not stop, 
because we would need to run it for an infinite amount of time, however we can decide if a program stops in $M$ steps.  
As far as there is a finite number of programs of length $K$ for each value of $K$ we can define a function $f(K)$ that 
is equal to the largest value of steps where a program of length $K$ stops.  All programs that do not stop in $f(K)$ 
must runs forever.  The previous construction could be done by checking the output of all the program of length less 
than $f(K)$.  Therefore if we could compute the function $f(K)$ or an upper bound to it, we would be able to get the 
list of all the program that stops.  

A simple analysis show that no contradiction is present if we the function $f(K)$ increases faster than any 
function that is computable with a program of constant length: e.g. we must have that for large $K$ that
\be
f(K)>e^{e^{e^{e^{e^{e^{e^{e^{e^{e^{e^{e^{K}}}}}}}}}}}}
\ee
The complexity of finding an upper bound to $f(K)$ must be $O(K)$.

The existence of a function that grows faster than what can be computed by a program of fixed length seems surprising 
however it is rather natural.  Indeed if we call $M(K)$ the maximum \emph{finite} number that can be printed by a 
program of length $K$.  Now it is evident that $M(K+1)>M(K)$, therefore the function $M(K)$ cannot be printed by any 
program with length less that $K$, so that there is no finite length program that is able to print all the values of 
$M(K)$.  The conclusion is that to find the shortest program (or if we prefer the shorted description) of a number $N$ 
is something that cannot be done by program: the complexity is a well defined quantity, but it cannot be computed 
systematically by a program.

Many variations on the same theme can be done: suppose that we add to our computer  an oracle: a specialized hardware 
that that tells the value of $f(K)$) (or directly the complexity of a string). We could now define a complexity 
$\Sigma_{1}(N)$ for the system \emph{computer +oracle}. The same argument as before tell us that the new complexity 
$\Sigma_{1}(N)$ cannot be computed for any $N$ by the the system \emph{computer +oracle}. We could introduce a 
superoracle and so on. The fact that we are unable, also with the use of any kind of oracles, to find a systems that can 
compute the computational complexity relative to itself recalls intuitively the Godel theorems: This is not a surprise 
as far there are deep relations among these arguments and  
with Godel theorems that  have been  investigated by Chaitin and presented also in many popular papers.

\section{The logical depth}
The problem of finding out the minimal complexity of a  number (or of a sequence of numbers), intuitively correspond to 
finding the rule according that some number have been generated.  This problem is  used in intelligence tests.  For 
example one may be asked to find out the rules that generate the following sequences or equivalently to predict the next 
numbers of the sequences
\begin{verbatim}
    1 2 3 4 5 6 7 8 9 10 11 12 13 
    1 2 4 8 16 32 64 128 256 512 1024 2048 
    1 1 2 3 5 8 13 21 34 55 89 144
    1 1 2 2 4 2 4 2 4 6 2 6 4 2 4 6 6 2 6 4 2 6 4 6 8 4 2 4
\end{verbatim}
For the first two lines  the rule is clear, the third line is a Fibonacci sequence: the next number is the sum of the of 
the two previous one, while the last line follow a slightly more complex rule and we leave to the reader the pleasure to 
find it. 

In the case that there are different rules, we consider a natural rule the simplest one: for example 
we could say that the first line is the sequence of natural integer that cannot be written as $a^{2}+b^{2}+c^{2}+d^{2}$ with 
 $a$,  $b$  $c$ and $d$ are all different (the first integer missing would be 30). However this second rule is much 
 longer and it is unnatural. The two rules would produce rather different sequences but if we know only the first 
 elements the fist choice is much more natural

A related and somewhat complementary concept is the logical depth of a number \cite{B}: roughly speaking it is the 
amount of CPU needed to compute it if we use the shortest program that generates it (i.e. the natural definition of the 
number).  

It is evident that if the complexity is given by the upper bound and the program consists of only one write instruction, 
the execution is very fast.  i.e. linear in the number of bits.  On the contrary, if a very short program prints a very 
long sequence, the execution may be still very fast, as in the case of a typical random number generator, or may be very 
slow, e.g. the computation of the first $10^6$ digits of $\pi$ or to find out the solution of an NP hard problem whose 
instance has been generated using a simple formula.  Only a low complexity sequence may have a large logical depth; 
moreover the same arguments of the previous section tell us that there must be for any value of the complexity $K$ 
sequences with extremely large values of the logical depth (i.e. of order $f(K)$.

These results concern us also because science aims to find out the simplest formulation of the law that reproduce the 
empirical data.  The problem of finding the simplest description of a complicated set of data correspond to find the 
scientific laws of the world.  For example both the Newton and the Maxwell laws summarize an enormous quantity of 
empirical data and are likely the shortest description of them.  The ability of finding the shortest description is 
something that cannot be done in general by a computer and and it is often taken a sign of intelligence.

Phenomenological explanations with lot of parameters and no so deep theory are often easy to apply and correspond to 
rules that have high complexity and low logical depth while simple explanations with few parameter and lot of 
computation needed have low complexity and rather high logical depths.  

For example the computation of chemical 
properties using the valence theory and the table of electronegativity of the various element belongs to the fists 
class, while a first principle computation, starting from basic formulae, i.e. quantum mechanics belongs to the second 
class.  A good scientific explanation is a low complexity theory (i.e. the minimal description of the data) that 
unfortunately may have an extremely large logical depth.  Sometimes this leads to the use of approximated theories with 
higher complexity and smaller logical depth.

\section{Learning form examples}
Scientific learning is just one case of a general phenomenon: we are able to lear from example and to classify the 
multitude of external object into different classes.  The problem of how it is possible to learn from examples has 
fascinated thinkers for a long time.  In the nutshell the difficulty is the following: if the rule cannot be logically 
derived from the examples \footnote{Two examples: (a) the horsiness is not the property of any given horse or set of 
horses (b) there are many different mathematical rules that generate the same sequence.} how do we find the rule?  The 
solution put forward by Plato was that the rule is already contained in the human brain and the examples have the only 
effect of selecting the good rule among all the admissible ones.

The opposite point of view (Aristotle) claims that the problem is ill posed and that human brain is 
tabula rasa before the experience of the external world.

Plato's point of view has been often dismissed as idealistic and non-scientific.  Here we want to suggest that this is 
not the case and that Platonic ideas are correct, at least in a slightly different contest.  Indeed the possibility of 
having machines which learn rules from examples has been the subject of intensive investigations in recent years 
\cite{AMIT}.  A very interesting question is to understand under which conditions the machine is able to generalize from 
examples, i.e. to learn the whole rule knowing only a few applications of the rule.  The main conclusion we want to 
reach is that a given machine cannot learn an arbitrary rule and the set of rules that can be easily be learned may be 
determined by analyzing the architecture of the machine.  Learning by example consists in selecting the correct rule 
among those contained in this set.  Let us see in the next sections how it may happen and how complexity is related to 
intelligence. 

\section{Learning, generalization and propensities}
In order to see how this selectionist principle may work let us start with some definition.  In 
the following I will consider rules which assign to each input vector of $N$ Boolean variables 
($\sigma_{i} = 0$ 
or 1 for $i=1,N$) an output which consists of a single Boolean value.  In other words a rule is a 
Boolean valued function defined on a set of $2^N$ elements (i.e. the set of all the possible values of 
the variables $\sigma$; it will be denoted by $R[\sigma]$).

The rule may be specified either by some analytic formulae or by explicitly stating the 
output for all the possible inputs.  The number of different rules increases very fast 
with $N$: it is given by $2^{2^{N}}\equiv 2^{N_{I}}$, where $N_{I}$ is the number of 
different possible input vectors, i.e. $2^{N}$.  In the following we will always consider 
N to be a large number: terms proportional to 1/N will be neglected.

A learning machine is fully specified if we know its architecture and the learning 
algorithm.  

Le us firstly define the architecture of the machine.  We suppose that the computations that the machine performs depend 
on M Boolean variables ($J_{k},\ k=1,M$).  In the nutshell the architecture is a Boolean function $A[\sigma,,J]$, which 
gives the response of machine to the input $\sigma$'s for each choice of the control parameters $J$'s.  Typical 
architectures are the perceptron \cite{MP} or a neural network, with discretized synaptic couplings.

For each given rule $R$ and choice of $J$'s, the machine may make some errors with respect to 
the rule $R$.
The total number of errors ($E[R,J]$) depends on the rule $R$ and on the $J$'s; it is given by
\begin{equation}
E[R,J]= \sum_{\{ \sigma \} } (R[\sigma]-A[\sigma,J])^{2} \ .   
\end{equation}

For a given architecture the machine may learn the rule $ R$ without errors if and only if 
is there exist a set of $ J$ 's such that $ E[R,J] = 0$ .  Simple counting arguments tell 
us that there are rules which cannot be learned without errors, if $ M$ is smaller that 
$2^{N}$ .  In most of the cases $ 2^{N}$ is extremely larger than $ M$ and therefore the 
number of admissible rules is an extraordinary tiny fraction of all the possible rules.

In a learning session we give to the machine the information on the values of $R[\sigma]$ for $L$ instances in the 
$\sigma$'s ($L$ is generally much smaller than $2^{N}$).  A learning algorithm tries to find the $J$'s which minimize 
the error on these $L$ instances \footnote{There are many different learning algorithm and some are faster than others.  
The choice of the learning algorithm is very important for practical purposes, but we will not investigate this point 
anymore.} Let us denote by $J^{*}$ the $J$'s found by the learning algorithm.

If the error on the other $2^{N-L}$ instances has decreased as effect of having learned the first $L$ 
instances, we say that the machine is able to generalize, at least to a certain extent.  Perfect 
generalization is achieved when no error is done on the other $2^{N-L}$ instances.

For a given machine the propensity to generalize depends on the rule and not all rules will be 
generalized by the machine.  Our aim is to understand how the propensity to learn different rules 
changes when we change the machine; in this note we are only interested in the effect of changing 
the architecture.

It was suggested by Carnevali and Patarnello in a remarkable paper \cite{CP} that, if we suppose that the 
learning algorithm is quite efficient, the propensity of the machine to generalize a given rule ($p_R$) 
depends only on the architecture.  The propensity ($p_R$) may be approximated by the number of 
different control parameters $J$ for which the total number of errors is zero.
	 
In other words we define the propensity as
\begin{equation}
p_R =2^{-M} \sum_{\{J\}} \delta ( E[R,J] ) \ ,   
\end{equation} 
where $E[R,J]$ is given by eq.  (1) and obviously depends only on the architecture. The function $\delta$ is defined is 
such a way that 
$\delta (k) = 1$ for $k = 0$\ , $\delta(k) = 0$ for $k \neq 0$ .

According to Carnevali and Patarnello, rules with very small pro\-pensity cannot be 
generalized, while rules with higher propensity will be easier to generalize.  In their 
approach the propensity of a given architecture in generalizing is summarized by the 
values of the function $p_R$ for all the $2^{N_{I}}$ arguments (the $ p_R$'s depend only 
on the architecture, not by the learning algorithms).  Of course the propensity cannot be 
directly related to the number of examples needed to learn a rule, a more detailed 
analysis, taking care of the relevance of the presented example, must be done.

\section{A statistical approach to propensities}
Our aim is to use statistical mechanics techniques \cite{P} to study in details the properties of $p_R$.

The $p_R$ are normalized to 1 ($\sum_{R} p_R = 1$) and it is natural to introduce the 
entropy of the algorithm $A$:
\begin{equation}
S[A]= - \sum_{R} p_R \ln(p_R).    \label{ENTROPIA}
\end{equation}
The entropy $S[A]$ is a non negative number smaller or equal than $\ln(2) \min(2^{N},M)$. 

We could say that if the entropy is finite for large $N$, the machine is able to represent 
essentially a finite number of rules, while, if the entropy is too large, too many rules 
are acceptable.

As an example we study the entropy of the perceptron (without hidden unity).  In this case 
(for $N =M $ odd) a detailed computation shows that all the $2^N$ choices of the $J'$s lead to 
different rules (i.e. two different $J$'s produce a different output at least in one case) 
and therefore $S[A] = \ln(2) N$.

We note that we could generalize the previous definition of entropy by introducing a 
partition function $Z(\beta)$ defined as follows
\begin{equation}
Z(\beta) = \sum_{R} \exp (\beta \ln(p_R)) =  \sum_{R} p_R^{\beta}.    
\end{equation}
We could introduce the entropy $S(\beta)$ associated with the partition function \cite{P} 
$(S(\beta)\equiv d{\ln[Z(\beta)/\beta]}/d\beta)$.  The previous defined entropy (eq. 
\ref{ENTROPIA}) coincides with $S(1)$.

The value of the entropy as function of $\beta$ tells us which is the the probability distribution of the 
$p_R'$s.  There are many unsolved question whose answer depend on the model: 
existence of phase transitions, structure of the states at low temperature, breaking of 
the replica symmetry \ldots

Many additional questions may be posed if we consider more than one architecture; in 
particular we would like to find out properties which distinguish between architectures 
which have similar entropies.  For example we could consider two different architectures 
(i.e. a layered perceptron or a symmetric neural network) with $N$ inputs and one output 
which have the same entropy (this can be achieved for example by adjusting the number of 
internal layers or hidden neurons).  It is natural to ask if these two different 
architectures are able to generalize the same rules, or if their propensity to generalize 
is concentrated on rules of quite different nature.  Our aim is to define a distance 
between two architectures, which will help us to compare the different performances.

Let us consider two architectures $A$ and $B$. A first step may consist in defining the 
entropy of $B$ relative to $A$ as
\begin{equation}
S[B/A]=  - \sum_{R} p_R(A) \ln[p_R(B)/p_R(A)] \ .    
\end{equation}
It can be shown that $S[B/A]$ is a non-negative quantity that becomes zero if and only if 
$p_R(A) = p_R(B)$.  The relative entropy is not symmetric and we can define the distance (or 
better the difference) between $A$ and $B $ as
\begin{equation}
d(A,B)= 1/2 {S[B/A] + S[A/B]}\ .    
\end{equation}
The introduction of a distance allow us to find out if two different architectures with 
the same generalization propensity; do they generalize the same rules or the rules they 
are able to generalize are different?  Unfortunately, the explicit computation of the 
distance among two architectures may be very long and difficult.

A very interesting and less understood question is how many examples are needed to  specify the rule for  a given 
architecture. The result obviously depend on the architecture (people with an high value of serendipity guess the write 
rule after a few examples) and explicit computations are very difficult, if we exclude rather simples cases.

\section{A possible definition of intelligence}
Having in our hands a definition of the distance between two architectures, we can now 
come to a more speculative question: how to define the intelligence of an architecture.  A 
possibility consists in defining an architecture $I(\sigma,J)$, which is the most 
intelligent by definition; the intelligence of $A$ can be defined as $-d[A,I]/S[A]$ (the 
factor $S[A] $has been introduced for normalization purposes).  The definition of the 
intelligent architecture $I$ is the real problem.

We suggest that a sequence of the most intelligent architectures is provided by a Turing 
machine (roughly speaking a general purpose computer) with infinite amount of time for the 
computation with a code of length $L$. More precisely the $J$'s are the $L$ bits of a code for 
the Turing machine (written in a given language) which uses the $\sigma$ as inputs.  The 
function $I(\sigma,J)$ is 1, if the program coded by $J$ stops after some time, and it is 0, if 
the the program never stops.  With this architecture we can compute the function $s(R)$  
(i.e. the simplicity of a rule) \cite{C3}, defined as $s(R)\equiv p_R(I)$.

It may be possible to estimate the function $s(R)$ using the relation $s(R) \approx 
2^{-\Sigma(R)}$, where $\Sigma(R)$ is the algorithmic complexity (introduced in the first section) i.e. the length of the 
shortest codes which computes the function $R[\sigma]$.

If we would like to know the intelligence of an architecture $A$ with entropy $S$, we should 
consider a Turing machine with nearly the same entropy and we should compute the distance 
between the two architectures.

The previous observations imply that algorithms which compute the rule in a relatively 
short time are likely unable to implement many rules with low algorithmic complexity and 
high logical depth.  In many of the most common algorithms the number of operations needed 
to compute the output for a given input is proportional to a power of $N$. For other 
algorithms (e.g. an asymmetric neural network, in which we require the computation of a 
limiting cycle) the computer time needed may be much larger (e.g. proportional to $2^L $).  It 
is natural to suppose that this last class of algorithms will be more intelligent than the 
previous one and will learn more easily rules with low algorithmic complexity and high 
logical depth \cite{PS}.

The puzzled reader may ask: if a general purpose computer is the most intelligent architecture, while people are studying 
and proposing for practical applications less intelligent architectures likes neural networks?  A possible answer is 
that this definition of intelligence may be useful to disembodied entities that have an unbound amount of time at their 
disposal.  If we have to find a rule and take a decision in real time (one second or one century, it does not matter, 
what matters is that the time is limited) rules of too low complexity and very large logical depth are completely 
useless; moreover a computer can be used to define the intelligence but we have seen that a computer is not well suited 
for finding and executing intelligent rules.  The requirement of taking a fast decision may be dramatic in some case, 
e.g. when you meet a lion on your path; evolution definitely prefer sa living donkey to a dead doctor and it quite 
likely that we have not been selected for learning rules with too large logical depth.

There are also other possible inconveniences with rules with too low complexity, i.e. they may be unstable with the 
respect of a small variation of the data they have to reproduce.  It is quite possible that the shortest program that 
pronounce words in Italian may have a completely different structure of the shortest program that pronounces words in 
Spanish, in spite of the similarity of the two languages. We would like to have a program that may be used for many 
languages, where we can add a new language without changing too much the core of the program; it is quite likely that such 
a program would be much longer than the shortest one for a given set of rules. Strong optimization and plasticity are 
complementary requirements. For example the nervous system of insects is 
extremely optimized and quite likely it is able to perform the tasks with a minimal number of neurons, but it certain lacks 
the plasticity of mammalian nervous system. 

Architectures that are different from a general purpose computer may realize  a compromise producing rules that have 
low, but not too low complexity and high, but not to high logical depth. Moreover these architectures are specialized: 
some of them (like neural networks) may be very good in working as associative memories but quite bad in doing arithmetics.
The problem  of finding the architecture that works in the most efficient way for a given task is difficult and 
fascinating and in many 
cases (e.g. reconstructing three dimensional objects from two dimensional images) it has also a very important practical 
impact.

I hope that this short exposition has convinced the reader of the correctness of the 
point of view that learning from example can be done only selecting among already 
existing rules. This is what typically happens in biology in many cases. The genetic information only preselect a large 
class of behavior and the external stimulus select the behavior among the available ones. This procedure can be seen 
in action in the immune systems. The number of possible different antibodies is extremely high (O$(10^{100}$)); each 
given individual at a given moment produces a much smaller number of different antibody (e.g. O($10^{8}$)). The external 
antigen select the most active antibodies among those present in the actual repertoire and stimulate their production. 
Eldeman has stressed that it is quite natural that a similar process happens also in the brain  as far as learning is 
concerned .

In conclusion at the present moment  we have only started our exploration of the 
properties of rules that are implicitly defined by an architecture.  I am convinced that 
the future will bring us many interesting results in this direction. It is amazing in how diverse directions 
Kolmogorov's ideas are relevant and find an application.

\include {kolmo.ref}

\end{document}